\documentclass[aps,prl,twocolumn,showpacs]{revtex4}
\usepackage{amssymb}
\usepackage{amsmath, amsthm}
\usepackage{color}
\usepackage{colordvi}
\usepackage{pstricks}
\usepackage{pst-plot}
\usepackage{graphicx}

\begin{document}


\title{Intra--Galactic thin shell wormhole and its stability}

\author{Ivana Bochicchio}
\email[]{ibochicchio@unisa.it}
\author{Ettore Laserra}
\email[]{ elaserra@unisa.it}
\affiliation{Dipartimento di Matematica, 
Universit\'a degli Studi di Salerno, Via Ponte Don Melillo, 84084 
Fisciano (SA), Italy}


\begin{abstract}
In this paper, we construct an intra-galactic thin shell wormhole joining two copies of identical galactic space times described by the Mannheim-Kazanas de Sitter solution in conformal gravity and study its stability under spherical perturbations. We assume the thin shell material as a Chaplygin gas and discuss in detail the values of the relevant parameters under which the wormhole is stable. We study the stability following the method by Eiroa and we also qualitatively analyze the dynamics through the method of Weierstrass. We find that the wormhole is generally unstable but there is a small interval in radius for which the wormhole is stable. 
\end{abstract}


\maketitle

\section{Introduction}

Wormholes are topological handles connecting two distant regions of spacetime. These objects are solutions of the Einstein
equations and have been known for a long time \cite{Ellis, staticwormholes}, but they are intensely investigated by physicists in recent times (see \cite{Visserbook} for an extensive discussion) after the seminal work by Morris and Thorne \cite{MorrisThorne} in 1988. We can mention only a few with the understanding that the list is by no means exhaustive. 
Dynamical wormholes were discovered and studied in Refs. \cite{HochbergVisser98, Hayward99, Lobo} and wormholes
in cosmological settings were contemplated in various works (\cite{Kim96, othercosmowormholes} and references therein), with
particular attention being paid to wormholes with cosmological constant $\Lambda$, which are asymptotically de Sitter or anti de Sitter according to the sign of  $\Lambda$ \cite{LemosLoboOliveira03}. 
These geometrical  constructions are designed to allow travelers to pass from one region of spacetime to another or even from one universe to another. It is also to be noted that wormholes are as valid solutions of Einstein's theory of gravity as are the black holes in the sense that no experiment has so far ruled out the existence of macroscopic wormhole structures in classical gravity. In the high energy experiments, such as LHC, microscopic wormholes are a distinct possibility.
On the other hand, it would be of interest to know if there could exist wormholes connecting two spacetimes of spherical symmetric galaxies. We are interested here in traversable wormholes but the situation is that, a humanly traversable wormhole solution covered by single regular coordinate chart is generally unavailable except in a few cases such as the Ellis wormhole \cite{Ellis} and in a particular class of solution in the Brans-Dicke theory \cite{Brans}. Thus, one has to use the cut and paste technology developed by Visser \cite{Visserbook, visser}, where one joins two copies of spacetimes across a throat threaded by a thin-shell of matter, to construct  a traversable wormhole. Some recent articles related to thin-shell wormholes can be found in Refs. \cite{altri,altri1,IvanaValerio}. Note that in \cite{IvanaValerio} a solution describing
a wormhole shell joining two identical Lemaitre-Tolman-Bondi (LTB) universes is analyzed (see \cite{ivanaLTB} and references therein for some remarks on these Universes). 
\par
The equation of state of the material making the shell is left to one's choice. We choose the equation of state of the Chaplygin gas because the free parameter in it can encompass different types of matter. The Chaplygin gas has recently considered by several authors (see, for example, Refs. \cite{altri1,chaplygin1} for a wormhole with this model for the matter and Refs. \cite{reason1,reason2} to deeply understand the reasons for its introduction).
A physically meaningful spherically symmetric galactic spacetime is the Mannheim-Kazanas de Sitter solution (MKdS) \cite{MKdS} in conformal gravity. This solution has the special feature that it does not require the hypothesis of dark matter in the galactic halo region, yet \textit{predicts} the observed flat rotation curves and also allows the computation of the sizes of galaxies \cite{MannheimBrien, nandi1} without requiring additional assumptions. We choose two copies of the MKdS solution for our construction of intra-galactic wormhole.
The paper is organized as follows: The next section details how to construct the thin-shell wormhole from spherically symmetric spacetime following the method by Eiroa \cite{Eiroa}, which include equations that should be satisfied by the throat radius and the energy conservation. Then we work out the expression for the potential energy in order to construct the Weierstrass equation. In Sect. 3, the formalism is applied to a particular spherically symmetric metric, namely, the MKdS metric. We assume that thin-shell matter is in the form of a Chaplygin gas. Under these conditions, we consider the equations that give the possible radii of the wormhole and determine their stability under spherical perturbations. In Sec 4., we present the general Weierstrass criterion and in Sec 5., we investigate the dynamical behavior of thin shell. Certain specific values of relevant parameters are also considered. Finally, in Sec. 6, the conclusions of this work are summarized. 
\section{Preliminary equations}
The spherically symmetric line element in polar coordinates $(t, r, \theta, \varphi)$ is 
\begin{equation}
\label{metric}
ds^2\,=\,-f(r) dt^2+f(r)^{-1}dr^2+h(r)(d\theta^2+\sin^2\theta d\varphi^2)
\end{equation}      
where $h(r)$ is always positive and $f(r)$ is a positive function for a given radius. 

Consider now a wormhole shell $\Sigma$ at $r=a$ which
joins two identical copies of the space time. Thin shell wormholes are not created naturally as the result of a dynamical process  after the big bang but are to be constructed by cut and paste method as follows. Thus we choose a radius $a$ greater than the event horizon radius $r_H$ (if the geometry \eqref{metric} has any) and cut two identical copies of the region with $r\geq a$ 
\begin{equation}\mathcal{M}^\pm\,=\,\{X^\alpha=(t,r,\theta,\varphi)/r\geq a\},
\end{equation}
and paste them at the two-dimensional hypersurface 
\begin{equation}
\Sigma \equiv \Sigma^\pm\,=\,\{X/F(r)\,=\,r-a=0\}.
\end{equation}
In this way, a new manifold $\mathcal{M}=\mathcal{M}^+ \cup \mathcal{M}^-$ is created with two mouths on each side. 
\\
If $h'(r)>0$ (condition of flare out), this construction creates a geodesically complete manifold representing a wormhole with two regions connected by a throat of radius $a$, where the surface of minimal area is located (see \cite{Eiroa}). Since the spacetime is not asymptotically flat, the resulting wormhole will have the topology of a dumbell.

In order to know the geometry on the thin shell connecting the two sides, we define on the shell coordinates $\xi^i\,=\,(\tau, \theta, \varphi)$, with $\tau$ the proper time on the shell. Moreover, using the Sen-Darmois-Israel formalism \cite{junction} and introducing the unit normal to $\Sigma$ in $\mathcal{M}$, it is possible to obtain the following expression for the second fundamental form (or extrinsic curvature) \cite{Eiroa}:
\begin{equation} K^\pm _{\widehat{\theta}\widehat{\theta}}\,=\,K^\pm _{\widehat{\varphi}\widehat{\varphi}}\,=\,\pm\frac{h'(a)}{2h(a)}\sqrt{f(a)+\dot{a}^2}\ ,\end{equation}
\begin{equation} K^\pm _{\widehat{\tau}\widehat{\tau}}\,=\,\pm\frac{f'(a)+2\ddot{a}}{2h\sqrt{f(a)+\dot{a}^2}}\ ,\end{equation}
where the orthonormal basis $\{e_{\widehat{\tau}}=e_{\tau}\,, e_{\widehat{\theta}}=a^{-1}e_{\theta}\,,e_{\widehat{\varphi}}=(a \sin \theta)^{-1}e_\varphi\}$ has been adopted and where a prime and the dot represent, respectively, the derivative with respect to $r$ and $\tau$.

Introducing the surface stress-energy tensor $S_{\widehat{i}\widehat{j}}=\,\,diag\,(\sigma,p_{\widehat{\theta}},p_{\widehat{\varphi}})$, 
where $\sigma$ is the surface energy density and $p_{\widehat{\theta}},p_{\widehat{\varphi}}$ 
are the transverse pressures, the Lanczos equations give \cite{Eiroa}:
\begin{equation} \label{densita}
\sigma\,=\,-\frac{\sqrt{f(a)+\dot{a}^2}}{4 \pi}\frac{h'(a)}{h(a)};
\end{equation}
and
\begin{equation} \label{pressione}
p=p_{\widehat{\theta}}=p_{\widehat{\varphi}}\,=\,\frac{\sqrt{f(a)+\dot{a}^2}}{8\pi}\left[	\frac{2\ddot{a}+f'(a)}{f(a)+\dot{a}^2}+\frac{h'(a)}{h(a)}\right].
\end{equation}
In particular, for \textit{static} wormholes with $a_0$ as radius, the surface energy density and the pressure are obtained by the previous equations putting $\dot{a_0}\,=\,0$.
\\
Moreover, the negative sign in \eqref{densita} plus the flare--out condition $h'(a)>0$ implies $\sigma<0$, indicating that the matter at the throat is ''exotic".
\\
In order to introduce the equation of state for the exotic matter at the throat, it is possible to suppose that such matter is a generalized Chaplygin gas. For this gas, the pressure has opposite sign to the energy density, resulting a positive pressure:
\begin{equation} \label{stato}
p\,=\,\frac{A}{|\sigma|^\alpha} \ ,
\end{equation}
where $A>0$ and $0<\alpha\leq 1$.When $\alpha\,=\,1$ the Chaplygin gas equation of state is recovered.
Replacing Eqs. \eqref{densita} and \eqref{pressione} in \eqref{stato}, it is possible to obtain the differential equation that should be satisfied by the throat radius of the thin shell wormhole (see \cite{Eiroa}). In particular, for \textbf{static} wormholes, it becomes:
\begin{equation}\label{equazionestatica}
\begin{array}{ll}
& \left[  f'(a_0)h(a_0)+f(a_0)h'(a_0)\right]\left[ h'(a_0)\right]^\alpha
\\[0.7em]
&-2A\left[ 4 \pi h(a_0)\right]^{\alpha+1}\left[ f(a_0)\right]^{(1-\alpha)/2}=0,
\end{array}
\end{equation}
with the condition $a_0>r_H$ if the original metric has horizons. 
\\
Moreover, from Eqs. \eqref{densita} and \eqref{pressione}, the energy conservation equation reads \cite{Eiroa}:
\begin{equation} \label{conservation}
\frac{d}{d\tau}(\sigma\mathcal{A})+p\frac{d\mathcal{A}}{d\tau}=\left\{[h'(a)]^2-2h(a)h''(a)\right\}\frac{\dot{a}\sqrt{f(a)+\dot{a}^2}}{2h(a)},
\end{equation}
where $\mathcal{A}\,=\,4 \pi h(a)$ is the area of the wormhole throat, $\frac{d}{d\tau}(\sigma\mathcal{A})$ the internal energy change of the throat and $p\frac{d\mathcal{A}}{d\tau}$ the work done by the internal forces of the throat; the r.h.s. represents the flux. Hence Eq. \eqref{conservation} can be written as (see \cite{Eiroa})
\begin{equation} \label{conservation2}
h(a)\sigma'+h'(a)(\sigma+p)+\left\{[h'(a)]^2-2h(a)h''(a)\right\}\frac{\sigma}{2h'(a)}=0.
\end{equation}
By the equation of state, $p$ is a function of $\sigma$, thus Eq. \eqref{conservation2} is a first order differential equation for which an unique solution with a given initial condition always exists. Then Eq. \eqref{conservation2} can be integrated to obtain $\sigma(a)$. Replacing  $\sigma(a)$ in Eq. \eqref{densita}, the dynamics of the wormhole throat is completely determined by the single equation
\begin{equation} \label{weie1}
\dot{a}^2\,=\,-V(a),
\end{equation}
with
\begin{equation} \label{pot.}
V(a)\,=\,f(a)-16\pi^2\left[\frac{h(a)}{h'(a)}\sigma(a)\right]^2.
\end{equation}
Equation \eqref{weie1} is the Weierstrass equation and it will analyzed in the Sects. 4 and 5 in order to study the {\em dynamics} of our constructed wormhole. Finally, in order to study the stability of the {\em static} wormhole one has to consider the second derivative of the potential.\footnote{By the results in \cite{Eiroa}, it is possible to consider a Taylor expansion of the potential energy up to the second order and observe that $V(a)\,=\,(1/2)V''(a_0)(a-a_0)^2+O[(a-a_0)^2]$, since $V(a_0)\,=\,V'(a_0)\,=\,0$.} Precisely, observing that the wormhole is stable under radial perturbation if and only if $V''(a_0)>0$, it will be useful recall the following expression \cite{Eiroa}:
\begin{equation} \label{derivata2}
\begin{array}{ll}
&V''(a_0) \,= \,
\\& 
f''(a_0)+\frac{(\alpha-1)[f'(a_0)]^2}{2f(a_0)}
+\left[\frac{(1-\alpha)h'(a_0)}{2h(a_0)}
+\frac{\alpha h''(a_0)}{h'(a_0)}\right]f'(a_0)
\\&
+(\alpha+1)\left[\frac{h''(a_0)}{h(a_0)}-\left(\frac{\alpha h'(a_0)}{h(a_0)}\right)^2\right]f(a_0).
\end{array}
\end{equation}
\par
The next section will be devoted to constructing the thin-shell wormhole in the MKdS spacetime. First, we will investigate the event horizons of the geometry in order to find the right interval of radial coordinate. Next, we will employ Eiroa's considerations using a recent new result on the maximal stable galactic size. Then, after the analysis of stability of static wormhole through the standard potential approach, we will exploit the Weierstrass method to classify also the dynamics of the MKdS thin wormhole. 

\section{Mannheim-Kazanas-de Sitter wormhole}

The spherically symmetric Mannheim-Kazanas-de Sitter (MKdS)  metric line element in polar coordinates $(t, r, \theta, \varphi)$ and in units G = 1, vacuum speed of light $c = 1$, is 
\begin{equation}
\begin{array}{ll}
ds^2=&-\displaystyle \left(1-\frac{2M}{r}+\gamma r -k r^2\right)dt^2
\\[0.7em]&
+\left(1-\frac{2M}{r}+\gamma r -k r^2\right)^{-1} dr^2 + r^2 d\Omega^2
\end{array}
\end{equation}
where M is the mass, $k\,(\equiv\,\frac{\Lambda}{3})$ is of the order of the cosmological constant and $\gamma>0$ is the conformal parameter in the (MKdS) solution of Weyl gravity .
\\
If the cosmological constant is positive and $k >\eta_1+ \eta_2$,
where 
\begin{equation}
\eta_1 = \frac{9 M \gamma +1}{54 M^2} \ , 
\end{equation}
\begin{equation}
\eta_2=\frac{\sqrt{108 M^2 (8 M \gamma +1) \gamma ^2+(9 M \gamma +1)^2}}{108 M^2}\ ,
\end{equation}
 the function
\begin{equation} \label{metrica}
f(r)\,=\,\displaystyle 1-\frac{2M}{r}+\gamma r -k r^2
\end{equation}
is always negative, so we take $ 0 < k < \eta_1+\eta_2 $. In this case the geometry has two horizons, which are placed at
\begin{equation}r^1_{_H}=2\sqrt{\frac{\gamma^2+3 k}{9 k^2}} \cos\frac{\theta+4\pi}{3}+\frac{\gamma}{3k}\ ,\end{equation}  
\\
\begin{equation}r^2_{_H}=2 \sqrt{\frac{\gamma^2+3 k}{9 k^2}} \cos\frac{\theta}{3}+\frac{\gamma}{3 k} \ ,\end{equation}
\\[.7em]
where $r^2_{_H}>r^1_{_H}$, 
\begin{equation}
\theta\,=\,\arctan(-2\sqrt{-\Delta_1}/q_1), 
\end{equation}
\begin{equation}
\Delta_1\,=\,-\frac{8 M \gamma ^3+\gamma ^2+36 k M \gamma -108 k^2 M^2+4 k}{108 k^4}\ ,
\end{equation}
\begin{equation}
q_1\,=\,\frac{-2 \gamma ^3-9 k \gamma +54 k^2 M}{27 k^3} \ .
\end{equation}
\begin{figure}[h]
\includegraphics[width=.49\textwidth]{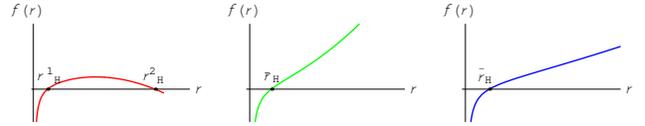}
\begin{minipage}{8cm}
\caption{ 
\small \it 
The three pictures represent the graph of the metric function $f(r)\,=\,1-\frac{2M}{r}+\gamma r - k r^2$ for different value of $k$. 
The red curve describes $f(r)$ when $ 0 < k < \eta_1+\eta_2$. The two horizons are marked by $r^1_{_H}$ and $r^2_{_H}$. 
The green curve represent the metric function for $k<\eta_1-\eta_2$. In such case the geometry has only one horizon placed at $\bar{r}_{_H}$. 
The blue curve is $f(r)$ for $0>k>\eta_1-\eta_2$. The unique horizon is marked by ${\tilde{r}}_{_H}$.
}
\end{minipage}
\end{figure}

When $k$ is negative and such that $k<\eta_1-\eta_2$ the event horizon is placed at 
\begin{equation}
\bar{r}_{_H}\,=\,\sqrt[3]{-\frac{q_2}{2}-\sqrt{\Delta_2 }}+\sqrt[3]{\sqrt{\Delta_2 }-\frac{q_2}{2}}-\frac{\gamma}{3|k|} \ .
\end{equation}
On the contrary, when $0>k>\eta_1-\eta_2$ the event horizon is placed at  
\begin{equation}\tilde{r}_{_H}\,=\,2\sqrt{\frac{\gamma^2-3|k|}{9 k^2}} \cos\frac{\theta+4\pi}{3}-\frac{\gamma}{3 |k|}
\end{equation}
where
\begin{equation}
\theta\,=\,\arctan\frac{-2\sqrt{-\Delta_2}}{q_2} \ , 
\end{equation}
\begin{equation}
\Delta_2\,=\,\frac{108 k^2 M^2+4 |k| (1+9M\gamma)-\gamma^2(1+8\gamma M)}{108 k^4} 
\end{equation}
and 
\begin{equation}
q_2\,=\,\frac{2 \gamma ^3-9 |k| \gamma -54 k^2 M}{27 k^3} \ .
\end{equation}
All these considerations are summarized in Fig. 1 
\par
The flare out condition holds ($h'(r)=2r>0$) in the constructed wormhole, hence the spacetime is a geodesically complete manifold with two regions connected by a throat of radius $a$, at which the surface of minimal area is located (see \cite{Eiroa}). Combining the idea of \cite{Eiroa} with the recent new results by Nandi and Bhadra \cite{nandi1} on the maximal stable limit $R^{^{max}}_{_{stable}}$ indicating the size of a galaxy, we take the radius $a$ greater than $R^{^{max}}_{_{stable}}$. This limit is computed from the generic condition for stability, viz.,
\begin{equation}
g(r)\,=\,2 f'^2(r)-f(r)f''(r)-3f(r)f'(r)/r < 0 \ .
\end{equation} 
The importance of this limit is that it should be regarded as the testable upper limit on the size of a galaxy. Observations on galactic sizes have so far respected this limit. The very fact that there exists a finite limit $R^{^{max}}_{_{stable}}$ distinguishes conformal theory from some dark matter models because in the latter there is no such limit. These facts distinguish the present thin shell wormhole from similar others.  

Hence, the surface of thin-shell exotic matter (the throat) should be limited within the radius range
\begin{equation}r^2_{_H}<R^{^{max}}_{_{stable}}<a_0<r^1_{_H}\end{equation}
if the cosmological constant is positive and $k > \eta_1+\eta_2$. 
\\
On the other hand, if the cosmological constant is negative and $k$ satisfy the condition $k<\eta_1-\eta_2$, the throat radius $a_0$ must satisfy
\begin{equation}
a_0>R^{^{max}}_{_{stable}}>\bar{r}_{_H}\ .
\end{equation}
Finally, when $0>k>\eta_1-\eta_2$ the radius is placed as follows
\begin{equation}a_0>R^{^{max}}_{_{stable}}>\tilde{r}_{_H}\ .\end{equation}
\par
By replacing Eq. \eqref{metrica} in Eqs. \eqref{densita} and \eqref{pressione}, and considering the static case,  the energy density and the pressure at  the throat become
\begin{equation}
\label{densita2}
\sigma_0 = -\frac{\sqrt{-k a_0^3 + \gamma a_0^2 + a_0 -2M}}{2 \pi a_0\sqrt{a_0}} \ ,
\end{equation}
and 
\begin{equation}
\label{pressione2}
p_0 = \frac{2 a_0-2M-4ka_0^3+3 \gamma a_0^2}{8 \pi a_0\sqrt{a_0}\sqrt{-k a_0^3 + \gamma a_0^2 + a_0 -2M}} \ .
\end{equation}
Moreover, from Eq. \eqref{equazionestatica} the throat radius $a_0$ should satisfy the equation
\begin{equation} \label{raggio}
\begin{array}{ll}
& 2^{\alpha } a_0^{\alpha } \left(-4 k a_0^3+3 \gamma  a_0^2+2
   a_0-2 M\right)
\\[0.7em]
&- 2^{2 \alpha +3} a_0^{2 (\alpha +1)} A \pi
   ^{\alpha +1} \left(-k a_0^2+\gamma  a_0+1-\frac{2
   M}{a_0}\right)^{\frac{1-\alpha }{2}}=0 \ .
\end{array}
\end{equation}
For a generalized Chaplygin gas, this equation can be solved numerically, while for $\alpha=1$ (Chaplygin gas) it is cubic in $a_0$, hence it can be solved analytically. In the latter case, one obtains:
\begin{equation}
\label{raggio2}
1- \frac{2\bar{M}}{a_0} + \bar{\gamma} a_0-\bar{k} a_0^2=0 \ ,
\end{equation}
where 
\begin{equation}\bar{M}\,=\,\frac{M}{2}\ , \quad \bar{\gamma}=\frac{3}{2}\gamma\ , \quad \bar{k}=2k+8 A \pi^2 \ .
\end{equation}
Following the above strategy to locate the horizons and substituting the values of the constants, we can find the solution of Eq. \eqref{raggio2} as different allowed values of the static throat radii. 
\par
In addition, in the MKdS metric, $[h'(a)]^2-2h(a)h''(a)=0$; hence the flux term in Eq. \eqref{conservation} is zero, so \eqref{conservation} takes the form of a conservation equation. Thus, in the case of Chaplygin gas $(\alpha=1)$, it becomes:
\begin{equation}
\label{differential}
\sigma'(a) a^2+2 \sigma(a) a-\frac{2 A a}{\sigma(a)}=0 \ .
\end{equation}
This equation can be easily integrated. As a general integral one obtains \footnote{ Only the negative solution is taken into account since we are considering exotic matter.}:
\begin{equation}
\label{soluzione}
\sigma(a)=-\sqrt{A+\frac{e^{2 C}}{a^4}}\ ,
\end{equation}
where $C$ is the integration constant. Now, replacing Eq. \eqref{soluzione} into Eq. \eqref{pot.}, we can write the explicit expression for the Weierstrass equation:
\begin{equation}
\label{weieprima}
\begin{array}{ll}
\dot{a}^2\,&=\,-f(a)+8 a A \pi^2 + 8 \pi^2 \frac{e^{2C}}{a^2}
\\[0.7em]&
= k a^2-\gamma  a-1+\frac{2 M}{a}+8 a A \pi^2 + 8 \pi^2 \frac{e^{2C}}{a^2} \ .
\end{array}
\end{equation}
For sake of convenience, let's choose the integration constant $C\,=\,-\frac{\log(8\pi^2)}{2}$ and the "constant of state'' $A\,=\,\frac{1}{8\pi^2}$. In such a way, Eq. \eqref{weieprima} becomes
\begin{equation}
\label{weieseconda}
\dot{a}^2 \, =\,  k a^2 + a (1-\gamma)-1+\frac{2M}{a}+\frac{1}{a^2} \  .
\end{equation}
\par
Finally, be replacing Eq. \eqref{metrica} in Eq. \eqref{derivata2} we obtain 
\begin{equation}
\label{potenziale}
\begin{array}{ll}
V''(a_0)=&-4 k +\frac{\gamma }{a_0}+\frac{(\alpha -1) \left(\frac{2 M}{a_0^2}-2 k a_0+\gamma
   \right)^2}{2 \left(-\frac{2 M}{a_0}+a_0 (\gamma -k
   a_0)+1\right)}
 \\[1em]  
& + \frac{2 (\alpha +1)
   \left(2 M+a_0 \left(k a_0^2-\gamma 
   a_0-1\right)\right)-2M}{a_0^3}
\end{array}
\end{equation}
The solutions of Eq. \eqref{raggio2} correspond to stable wormholes if $V''(a_0)>0$. 

\subsection{Chaplygin gas as the exotic matter on the shell $\Sigma$: the analysis of stability}
In order to analyze the stability of the constructed wormhole we adopt the Chaplygin gas ($\alpha=1$) as the exotic matter in the shell $\Sigma$. Hence Eq. \eqref{potenziale} becomes:
\begin{equation}
\label{potenziale1}
V''=\frac{6 M- a_0 (3 a_0 \gamma +4)}{a_0^3}.
\end{equation}
In this case, we observe that the stability analysis is completely independent of the state constant $A$ as well as the solution constant $k$. This means that for de Sitter or anti de Sitter Universe and for fixed value of $\gamma$ and $M$ we construct the same ``stable interval'':
\begin{equation}a_0 \in \left(0, \bar{a}_0\,=\, \frac{-2+\sqrt{4+18 \gamma M}}{3 \gamma} \right) \ .\end{equation}
Now we have to compare the admissible values of $r$ for $f(r)$ with the stable interval. In such way the stability problem is completely solved as follows:
\begin{figure}[t]\label{potenziali1}
\centering
\includegraphics[width=.52\textwidth]{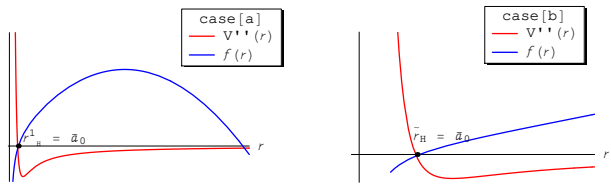}
\begin{minipage}{8cm}
\caption{ \small \it 
In these figures the relationship between $V''$ (see Eq. \eqref{potenziale1}) and $f(r)$ is underlined. On the left, $V''$ and $f$ are plotted for the admissible positive values of $k$. In this case the radii which make $f(r)\geq0$ lead to $V''$ being negative. This means instability for the static wormhole shell. On the right, the two functions are plotted for negative k but greater than the value $\eta_1-\eta_2$. Also in this case the second derivative of the potential is negative so that the wormhole shell is unstable. }
\end{minipage}
\end{figure}

\begin{figure}[h]\label{caso3}
\centering
\includegraphics[width=.4\textwidth]{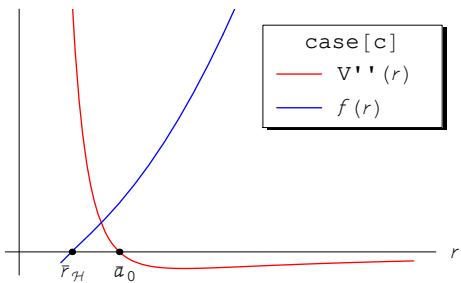}
\begin{minipage}{8cm}
\caption{ \small \it 
In this figure the relationship between $V''$ (see Eq. \eqref{potenziale1}) and $f(r)$ is underlined for negative $k$ but less than the value $\eta_1-\eta_2$. In this case the admissible values of the static radius are divided into two interval: the smallest one of stability and the greatest one of instability. When $\bar{r}_H \leq a_0 \leq \bar{a}_0$, $V''$ is positive: the wormhole is stable; for all the other values $a_0\geq \bar{a}_0$, $V''$ changes sign and the wormhole becomes unstable. }
\end{minipage}
\end{figure}

\begin{enumerate}
	\item [a.] When $ 0 < k < \eta_1+\eta_2$, $f(r)$ is positive in the interval between the horizons, but the function $V''$ is negative for all $a_0 \in (r_{_H}^1, +\infty)$ (see Fig. 2a.): the shell--wormhole is unstable.
	\item [b.] When  $0>k>\eta_1-\eta_2$, $f(r)$ is positive for $r$ greater than the horizon, but the function $V''$ is negative for all $a_0 \in (\tilde{r}_{_H}, +\infty)$ (see Fig. 2b.): also in this case the shell-wormhole is unstable. 
\item [c.] Finally, when $k$ is negative and such that $k<\eta_1-\eta_2$, $f(r)$ is positive for $r$ greater than the horizon and the behavior of the wormhole is very interesting (see Fig.3). In fact, contrary to the previous two cases, there is a small interval where it is \textit{stable}. When $\bar{r}_H \leq a_0 \leq \bar{a}_0$ the wormhole is stable; for all the other values $a_0\geq \bar{a}_0$ it becomes unstable. The size of the interval $(\bar{r}_H, \bar{a}_0)$ is strictly related to $k$ and it increases with the decreasing of $k$.
\end{enumerate}

\section{Generalization of the Weierstrass criterion to the  MKdS--wormhole}
\subsection{The Weierstrass equation}
In this section we will present a generalized form of the classical Weierstrass criterion,\,%
which can be applied to many problems outside of classical mechanics too \cite{ivana}. 
\\
Let's consider the first order differential equation \begin{equation}\label{eq:Weierstrass}
\dot{a}^{2} = \Phi({a})\ .
\end{equation}
Eq. \eqref{eq:Weierstrass} is called \textit{Weierstrass equation} and the function $\Phi({a})$ is called \textit{Weierstrass function}. The Eq. \eqref{eq:Weierstrass} can be written as
\begin{equation}\label{eq:sqrtWeierstrass}
\frac{da}{d\tau} = \pm\sqrt{\Phi({a})}\ ,
\end{equation}
which can be integrated by separating the variables
\begin{equation} \label{eq:intGenWeie}
\tau(a) = \pm\int_{a_0}^x\frac{ da}{\sqrt{\Phi(a)}} + \tau_0\ ,
\end{equation}
where we choose the sign $\pm$ in agreement with the sign of the initial rate $\dot{a}_0$.
\\
The importance of the Weierstrass approach is mainly based on the
fact that it is possible to obtain the qualitative behavior of the solutions of a Weierstrass
equation without integrating it. In such qualitative analysis the zeros of the
Weierstrass function have a leading role: the solutions of the Weierstrass equation are confined in
those regions of the ${a}$--axis where $\dot{a}^2\geq 0$. These regions are defined by the zeros of the Weierstrass function. These zeros are called \textit{barriers}, because they cannot be crossed by the solution $a(\tau)$.
The barriers split the range of possible values for $a$
into \textit{allowed} (in which $\Phi(a) \geq 0$) and \textit{prohibited} intervals (in which $\Phi(a) \leq 0$).
If a barrier $a_B$ is a simple zero of $\Phi$ such that
\begin{equation}
\Phi({a}_{B}) = 0\ ,\quad\Phi'({a}_{B})\neq 0\ ,
\end{equation}
then it is called an \textit{inversion point}
because the motion reverses its course after reaching it.
If a barrier ${a}_{B}$ is a multiple zero such that
\begin{equation}
\Phi({a}_{B})=0\ ,\quad\Phi'({a}_{B}) = 0\ ,
\end{equation}
then ${a}_{B}$ separates two allowed intervals
and it is called a \textit{soft barrier}.
A soft barrier is an \textit{asymptotic position},
because it is reached in an infinite time. In fact, for $a={a}_{B}$ the integral in Eq. \eqref{eq:intGenWeie} becomes diverging.
Finally we recall that an asymptotic position is also an equilibrium position. This means that if at the initial time $\tau_0$, $a_0=a(\tau_0)\,=\,{a}_{B}$, then $a(\tau)\,=\,{a}_{B} \,\, \forall\,\,\tau$.
So, once these zeros are found, the qualitative behavior of the
solutions of the Weierstrass equation \eqref{eq:Weierstrass} is
completely determined.

\subsection{The Weierstrass criterion for the evolving MKdS wormhole}
In order to qualitatively study the behavior of dynamic MKdS wormhole, let us analyze Eq.  \eqref{weieseconda}
which determines the evolution of the dynamic wormhole through the Weierstrass method. We let $a \rightarrow a(\tau)$ and consider the quadratic differential equation \eqref{weieseconda}
\begin{equation} \label{weie2-ripetuto}
\begin{array}{ll}
\dot{a}^{2}=-V(a)&=-f(a)+16\pi^2\left[\frac{h(a)}{h'(a)}\sigma(a)\right]^2
\\[0.7em]&= k a^2+(1-\gamma)a-1+\frac{2M}{a}+\frac{1}{a^2}
\ ,
\end{array}
\end{equation}
that is a Weierstrass equation with Weierstrass function
\begin{equation} \label{eq:WeierstrassPhi}
\Phi({a}) = k a^2+(1-\gamma)a-1+\frac{2M}{a}+\frac{1}{a^2}
\end{equation} 
depending on the parameter $a$. Equation
\eqref{weie2-ripetuto} translates into two equations
\begin{equation}
\begin{array}{ll}
\frac{d{a}}{d \tau}
&= \pm\sqrt{\Phi({a})}
\\&
= \pm\sqrt{k a^2+(1-\gamma)a-1+\frac{2M}{a}+\frac{1}{a^2}}
\ ,
\end{array}
\end{equation}
where we have to choose the sign $\pm$ in agreement with the sign of the initial rate $\dot{a}_0$.
The zeros of the Weierstrass function \eqref{eq:WeierstrassPhi} depend on the sign of
$k$ (and hence of $\Lambda$) and on the fixed value of the conformal parameter $\gamma$. These different
situations will be analyzed in detail in the following sections,
where the zeros of the Weierstrass function are obtained by
finding the positive real roots of the fourth degree equation
\begin{equation}\label{terzo_grado}
k a^4 + (1-\gamma)a^3-a^2+2Ma+1=0 \ .
\end{equation}
Note that since
\begin{equation}\Phi(a)\,=\,-f(a)+a+\frac{1}{a},
\end{equation}
if we evaluate $f$ in the zeros of the Weierstrass function, we obtain positive values. This is in agreement with the geometry.
\\
\section{Zeros of the Weierstrass function and dynamic behavior of the wormhole.} 

\subsection{Conformal parameter $\gamma \leq 1$} The positivity of the coefficient of $a$ in Eq. \eqref{eq:WeierstrassPhi} implies a forever expanding wormhole for
positive $k$, while a different and more complex situation is obtained for negative $k$.

When $k$ is positive, the Weierstrass function has no real and positive zeros.
Hence, having no barriers, the expansion of the wormhole is monotonic: if the MKdS wormhole is initially
expanding, it will go on expanding without limit. On the contrary, when $k$ is negative the Weierstrass
function \eqref{eq:WeierstrassPhi} has a simple zero $a_M$. Since simple zeros are inversion points, the wormhole can expand until its throat radius reaches the maximum expansion radius given by $a_M$. After reaching this value, the radius of wormhole throat decreases. Thus, in a finite time the wormhole could collapse. These two cases are represented in Fig. 4.

\begin{figure}[t]
\centering
\includegraphics[width=0.5\textwidth]{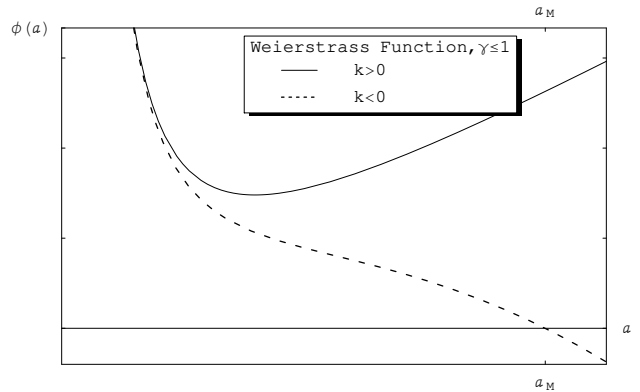}
\begin{minipage}{8cm}
\caption{ \small \it 
The Weierstrass function $\Phi({a})$ for
$\gamma \leq 1$ in two different cases:
i) when $k>0$ there are no barriers: the wormhole expansion is possible for all $a(\tau)$ belonging to the admissible values for $a$;
ii) when $k<0$ the function $\Phi(a)$ has a simple zero $ {a}_M$. Since it is an inversion point the expansion is possible $ \forall a \in (0, {a}_M]$ 
}
\end{minipage}
\end{figure}
\subsection{Conformal parameter $\gamma> 1$}
For negative value of the coefficient of $a$  in Eq. \eqref{eq:WeierstrassPhi}, the features of the Weierstrass function depend on the sign of $k$. For positive value of $k$, three different situations can occur: $\Phi(a)$ can have no zeros, simple zeros or multiple zeros. On the other hand, negative value of $k$ implies the existence of a soft barrier, hence the behavior of the wormhole strongly depends on the initial conditions.
\begin{figure}[t]
\includegraphics[width=0.5\textwidth]{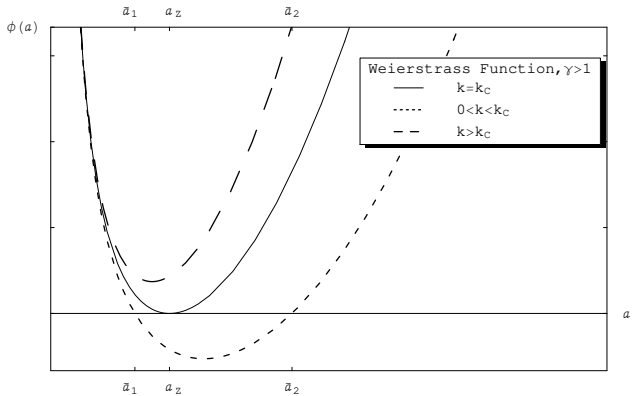}
\begin{minipage}{8cm}
\caption{ \small \it The Weierstrass function $\Phi({a})$ for
$\gamma >1$ and positive $k$. The behavior of $\Phi({a})$ depends on $k$:
i) when $k = k_c$, $\Phi({a})$ has a multiple zero at ${a}_{_Z}$. An equilibrium position or a limiting
position corresponds to it. In the first case,
$a\,=\,{a}_{_Z}$ is the unique possible position. In the second
case, the expansion is possible $\forall \,\, 0 < {a} < {a}_{_Z}$ and  $\forall \,\,  {a} > {a}_{_Z}$ but the respective behaviors of the wormhole are different;
ii) when $0<k < k_c$, the function $\Phi({a})$ has two
simple zeros at $\bar{a}_{1}$ and $\bar{a}_{2}$ (i.e. there are
two inversion points $\bar{a}_{1}$ and $\bar{a}_{2}$). The expansion
is possible $\forall \,\, 0 < {a} \leq \bar{a}_{1}$ and $\forall
a\geq \bar{a}_{2}$;
iii) when $k > k_c$ the Weierstrass function has no
zeros. The expansion $\forall\, a > 0$ is possible.}
\end{minipage}
\end{figure}

When $k$ is positive it is possible to find a critical value $k_c$ such that when $k=k_c$, the Weierstrass function has one multiple zero $a_Z$. In this situation the behavior of the MKdS wormhole depends on the initial
conditions: its evolution is static if the initial
radius $a_0$ equals $a_Z$. On the other hand, the static
situation is a limiting situation: if $a_0\,\neq\,a_Z$ and $a_0\,<\,a_Z$, the
wormhole goes on expanding asymptotically approaching the static
model with $a_Z$ as radius. On the contrary, for $a_0\,>\,a_Z$ the wormhole expands without limit. 
\\
If $k\,>\,k_c$, there are no barriers, hence if the MKdS wormhole is initially expanding $(\dot{a}_0>0)$,
it will go on expanding without limit.
\\
Finally, when $0<k<k_c$, the Weierstrass function \eqref{eq:WeierstrassPhi}
admits two simple zeros that can be called $\bar{a}_1$ and $\bar{a}_2$.
So when the initial radius of the wormhole is less that $\bar{a}_1$ and
the wormhole is initially expanding, it will go on expanding
until its radius reaches the maximal expansion value
$\bar{a}_1$; then it will contract back from $\bar{a}_1$. Hence it could collapse in a finite time. On
the other hand, if the MKdS wormhole is expanding from the initial radius greater than
$\bar{a}_2$, it will
go on expanding. Once more, if $a_0>\bar{a}_2$ and if the expansion velocity is negative, the wormhole is contracting and its radius will continue to decrease until it reaches the value $\bar{a}_2$. Since $\bar{a}_2$ is an inversion position, the wormhole will start to grow and it will continue to grow indefinitely. Note that the evolution of the wormhole, either to indefinite expansion or collapse, depends strongly on the initial conditions. Moreover, a peculiar situation is observed when the initial radius $a_{0}$ is such that $\bar{a}_1 < a_{0} <
\bar{a}_2$, since in this case the Weierstrass function is negative: the initial radius of the wormhole can't belong to the
interval $(\bar{a}_1, \bar{a}_2)$. These results are summarized in Fig. 5.
\par
The last case is related to negative values of $k$. 
In this case, the zero of the Weierstrass function $a_{_I}$ is a simple one, i.e. an inversion position. Hence, we have $\Phi(a)\leq 0$ for $a \geq a_{_I}$: the evolution is possible for all values $0\leq a\leq a_{_I}$ .
When $\gamma\,>\,1$ and $k<0$, the wormhole is closed, that is,
if the wormhole shell is initially expanding, it will continue to expand until the radius reaches its maximum value $a_{_I}$; then it will contract back from $a_{_I}$ to eventually collapse in a finite time. This case is represented in Fig. 6.

\begin{figure}[h]
\includegraphics[width=0.5\textwidth]{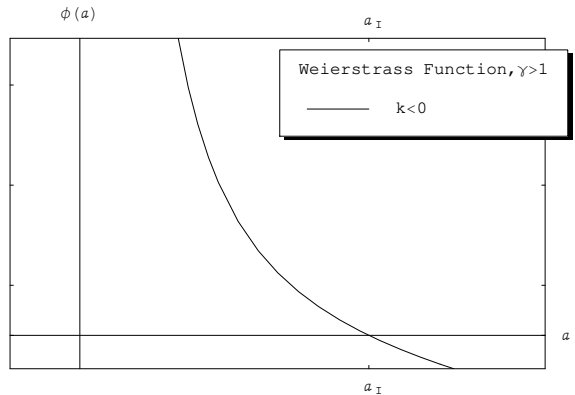}
\begin{minipage}{8cm}
\caption{ \small \it The Weierstrass function $\Phi({a})$ for negative
$k$ and $\gamma>1$. In this case the function $\Phi({a})$ has a simple
zero at ${a}_{_I}$ (i.e. there is one inversion point ${a}_{_I}$)
and the expansion $\forall \,\, 0 < {a} \leq {a}_{_I}$ is possible.}
\end{minipage}
\label{fig6}
\end{figure}

\section{Conclusions}

We have constructed an intra-galactic thin shell wormhole joining two copies of identical galactic space times described by the Mannheim-Kazanas-de Sitter solution in conformal gravity and interpreted the wormhole behavior using analytical methods. The approaches we have used are the Eiroa's potential approach for stability and the Weierstrass method for the dynamic evolution. It was shown that the wormhole is generally unstable but there is a small interval in radius for which the wormhole \textit{is} stable depending on the different signs of the cosmological constant.

The MKdS solution contains two arbitrary parameters, $\gamma$ and $k (\equiv \frac{\Lambda}{3})$ that
are known to play prominent roles in the galactic halo. In particular, $k$ provides the global quadratic potential that is essentially needed to explain the observed flat rotation curves \cite{MannheimBrien,nandi1}. In the present paper, we have shown how this constant can determine the stability and also the dynamical behavior of the thin-shell wormhole.

Our analysis included, among other mathematically allowed values, also the well accepted physical values of $k$ and $\gamma$, which are
$$k\,=\, 9.54 \times 10^{-54} cm^{-2} \ , \quad \gamma\,=\,3.06 \times 10^{-30}  cm^{-1}  \ .$$
These values consistently predict flat rotation curves without the hypothesis of dark matter.
For the mass of the galaxies, we considered the typical value $M\,\approx \,10^{16}$. With these data we were able to characterize the stability of the wormhole and predict their evolution. In particular
\begin{equation}
\eta_1-\eta_2=\,9.26 \times 10^{-35} > k >0
\end{equation}
which, as can be seen, is an admissible physical value because it falls within the range above. For this value of $k$, one has two horizons, as explained in Sect. 3 and in this case the static wormhole is unstable while from the dynamical point of view it is an ever expanding thin-shell wormhole. These results suggest that a real travel through wormhole from one galaxy to another is nearly impossible.

\section{Acknowledgments}

The authors are extremely grateful to Prof. Kamal Kanti Nandi for encouragement and discussion.


\begin{thebibliography}{99}


\bibitem{Ellis}
H.G. Ellis, {\em J. Math. Phys.} {\bf
14}, 104 (1973); {\em Gen. Relat. Gravit.} {\bf 10}, 105 (1979).

\bibitem{staticwormholes} 
K.A. Bronnikov, {\em Acta Phys. Pol. B} {\bf 4}, 251 (1973); T.
Kodama, {\em Phys. Rev. D} {\bf 18}, 3529 (1978); G. Cl\`ement,
{\em Gen. Relat. Gravit.} {\bf 13}, 763 (1981); D.H. Coule and K.
Maeda, {\em Class. Quantum Grav.} {\bf 7}, 955 (1990); D.H. Coule,
{\em Class. Quantum Grav.} {\bf 9}, 2353 (1992);  C. Barcelo and
M. Visser, {\em Phys. Lett. B} {\bf 466}, 127 (1999); C.
Armendariz-Picon, {\em Phys. Rev. D} {\bf 65}, 104010 (2002).


\bibitem{Visserbook} M. Visser, {\em Lorentzian Wormholes: From
Einstein to Hawking} (AIP Series in Computational and Applied
Mathematical Physics, Woodbury, USA, 1995).

\bibitem{MorrisThorne} M.S. Morris and K.S. Thorne, {\em Am. J. Phys.}
{\bf 56}, 395 (1988).


\bibitem{HochbergVisser98} D. Hochberg and M. Visser, {\em Phys.
Rev. D} {\bf 58}, 044021 (1998).

\bibitem{Hayward99} S.A. Hayward, {\em Int. J. Mod. Phys. D} {\bf
8}, 373 (1999).

\bibitem{Lobo}
F.S.N. Lobo and P. Crawford, {\em Class. Quantum Grav.} {\bf 22}, 4869 (2005).


\bibitem{Kim96} S.-W. Kim, {\em Phys. Rev. D} {\bf 53}, 6889
(1996).

\bibitem{othercosmowormholes} D. Hochberg and T.
Kephart, {\em Gen. Relat. Gravit.} {\bf 26}, 219 (1994); M.S.R.
Delgaty and R.B. Mann, {\em  Int. J. Mod. Phys. D} {\bf 4}, 231
(1995); D. Hochberg, {\em Phys. Rev. D} {\bf 52}, 6846 (1995); C.
Barcelo, {\em Int. J. Mod. Phys. D} {\bf 8}, 325 (1999); E.F. Eiroa and C. Simeone, {\em Phys. Rev. D} {\bf 71}, 127501 (2005);
E.F. Eiroa, {\em Phys. Rev. D} {\bf 78}, 024018 (2008);
S.V. Sushkov and Y.-Z. Zhang, {\em Phys. Rev. D} {\bf 77},
024042 (2008); E. Ebrahimi  and N. Riazi, {\em Astrophys. Sp.
Sci.} {\bf 321}, 217; M.  Cataldo, S. del Campo, P. Minning, and
P. Salgado, {\em Phys. Rev. D} {bf 79}, 024005 (2009).


\bibitem{LemosLoboOliveira03} J.P.S. Lemos, F.S.N. Lobo and S.
Q. de Oliveira, {\em Phys. Rev. D} {\bf 68}, 064004 (2003); F.S.N. Lobo and P.
Crawford, {\em Class. Quantum Grav.} {\bf 21}, 391 (2004).


\bibitem{Brans}
L.A. Anchordoqui, S.P. Bergliaffa and D. F. Torres, {\em Phys. Rev. D} {\bf  55}, 5226 (1997); K.K. Nandi, B. Bhattacharya, S.M.K. Alam, J. Evans, {\em Phys.Rev. D} {\bf 57}, 823 (1998);  A. Bhadra, K. Sarkar, D.P. Datta, K.K. Nandi, {\em Mod.Phys.Lett. A} {\bf 22} 367 (2007);  E. F. Eiroa, M. G. Richarte, C. Simeone, {\em Phys.Lett. A} {\bf 373} (2008);
A. Bhattacharya, I. Nigmatzyanov, R. Izmailov, K. K. Nandi,
{\em  Class.Quant.Grav.} {\bf 26}, 235017 (2009); A. Bhattacharya, R. Izmailov, K. K. Nandi {\em Class. Quantum Gravity} {\bf 26} (2010); F.S.N. Lobo, M.A. Oliveira, {\em Phys. Rev. D} {\bf 81}, 067501 (2010); N. M. Garcia, Francisco S.N. Lobo, {\em Mod.Phys.Lett.A} {\bf 40}, 3067 (2011); A Bhattacharya, R. Izmailov, E. Laserra, K. K. Nandi, {\em Class. Quantum Gravity} {\bf 28} (2011).

\bibitem{visser}
M. Visser, {\em Phys. Rev. D} {\bf 39} 3182 (1989); {\em Nucl.
Phys. B} {\bf 328} 203 (1989); E. Poisson and M. Visser, {\em Phys. Rev. D} {\bf 52} 7318
(1995).



\bibitem{altri}
M.G. Richarte, C. Simeone, {\em Phys. Rev. D} {\bf 76}, 087502 (2007); \textit{ibid.} {\bf 77},089903 (E) (2008); 
 K.A. Bronnik, A.A. Starobinsky, {\em Mod. Phys. Lett. A} {\bf 24}, 1559 (2009); M. Ishak and K. Lake, {\em Phys. Rev. D} {\bf 65}, 044011 (2002).


\bibitem{altri1}
E.F. Eiroa, C. Simeone, {\em Phys. Lett. A} {\bf 373}, 1 (2008);  \textit{ibid.} {\bf 373}, 2399 (E) (2009).

\bibitem{IvanaValerio}
I. Bochicchio, V. Faraoni, {\em Phys. Rev. D} {\bf 82}, 044040 (2010).

\bibitem{ivanaLTB}
I. Bochicchio, E. Laserra, {\em ICCS 2007, Part II} {\bf Lecture Notes in Computer Science} {\bf 4488 LNCS (PART 2)}, 997 (2007); I. Bochicchio, E. Laserra, {\em Journal of Interdisciplinary Mathematics} {\bf 12}, 537 (2009); I. Bochicchio, M. Francaviglia, E. Laserra, {\em Int. J. Geom. Methods Mod. Phys.} {\bf 6},  595 (2009), I. Bochicchio, V. Faraoni, {\em Gen. Relat. Grav.} {\bf 44}, 1479 (2012), doi: 10.1007/s10714-012-1350-7.

\bibitem{chaplygin1}
A. Kamenshchik, U. Moschella, V. Pasquier, {\em Phys. Lett. B} {\bf 511}, 265 (2001); M.C. Bento, O. Bertolami and A.A.
Sen, {\em  Phys. Rev. D} {\bf 66}, 043507 (2002); V. Gorini, A. Kamenshchik, U. Moschella
and V. Pasquier, {\em gr-qc/0403062}; V. Gorini, U. Moschella, A.Yu. Kamenshchik, V. Pasquier, and A.A. Starobinsky, {\em Phys. Rev. D} {\bf 78}, 064064 (2008); S. Chakraborty and T. Bandyopadhyay, Int. J. Mod. Phys.D 18, 463 (2009); M. Jamil,
M. U. Farooq and M. A. Rashid, Eur. Phys. J. C 59, 907 (2009).

\bibitem{reason1}
S. Chaplygin, {\em Sci. Mem. Moscow Univ. Math. Phys.} {\bf 21}, 1 (1904); H.-S-Tien, {\em J. Aeron. Sci.} {\bf 6},
399 (1939); T. von Karman, {\em J. Aeron. Sci.} {\bf 8}, 337 (1941).

\bibitem{reason2}
V. Sahni and A. A. Starobinsky, {\em Int. J. Mod. Phys. D} {\bf 9}, 373 (2000); P. J. Peebles and B.
Ratra, {\em Rev. Mod. Phys.} {\bf 75}, 559 (2003); T. Padmanabhan, {\em Phys. Rep.} {\bf 380}, 235 (2003); P. F. Gonzalez-Diaz, {\em Phys. Rev. Lett.} {\bf 93}, 071301 (2004); S. V. Sushkov, {\em Phys. Rev. D} {\bf 71}, 043520 (2005); V. Faraoni and W. Israel, {\em Phys. Rev. D} {\bf 71}, 064017 (2005); F. S. N. Lobo, {\em Phys. Rev. D} {\bf 71}, 084011 (2005); {\em ibid.} {\bf 71}, 124022 (2005); K. A. Bronnikov and A. A. Starobinsky, {\em JETP Lett.} {\bf 85}, 1 (2007); A. DeBenedictis, R. Garattini and F.S.N. Lobo, {\em Phys. Rev. D} {\bf 78}, 104003 (2008); M. Cataldo, P. Labrana, S. del Campo, J. Crisostomo and P. Salgado, {\em Phys. Rev. D} {\bf 78}, 104006 (2008); J. A. Gonzalez, F. S. Guzman, N. Montelongo-Garcia, and T. Zannias,  {\em Phys. Rev. D} {\bf 79}, 064027 (2009).


\bibitem{MKdS} P. D. Mannheim and D. Kazanas, {\em Astrophys. J.} {\bf 342}, 635 (1989); P. D.
Mannheim, {\em Astrophys. J.} {\bf 479}, 659 (1997); P. D. Mannheim, {\em Phys. Rev. D}
{\bf 75}, 124006 (2007). See also the review: P.D. Mannheim, {\em Prog. Part. and Nucl.
Phys.} {\bf 56}, 340 (2006).

\bibitem{MannheimBrien}
P.D. Mannheim and J.G. O'Brien, {\em Phys. Rev. Lett.} {\bf 106},
121101 (2011).

\bibitem{nandi1}
K. K. Nandi and A. Bhadra, {\em Phys. Rev. Lett.} {\bf 109}, 079001 (2012). 

\bibitem{Eiroa} E.P.Eiroa, {\em Phys. Rev. D} {\bf 80}, 044033 (2009). 

\bibitem{junction} 
N. Sen, {\em Ann. Phys. (Leipzig)} {\bf 73}, 365 (1924); C. Lanczos, {\em Ann. Phys. (Leipzig)} {\bf
24}, 518 (1924), G. Darmois, {\em Memorial des Sciences Mathematiques} Fascicule {\bf XXV}, Chap. V (Gauthier-Villars, Paris, 1927); W. Israel, {\em Nuovo Cimento} {\bf 44B}, 1 (1966); {\bf 48B}, 463(E) (1967); P.Musgrave and K. Lake, {\em Class. Quantum Grav.}  {\bf 13}, 1885 (1996).

\bibitem{ivana}
I. Bochicchio, E. Laserra, {\em Journal of Interdisciplinary Mathematics} {\bf 10}, 747 (2007), Bochicchio, E. Laserra,  {\em Gen. Relativ. Grav.}, {\bf 41} 2813 (2009), doi:10.1007/s10714-009-0809-7; I. Bochicchio, S. Capozziello, E. Laserra, {\em Int. J. Geom. Methods Mod. Phys.} {\bf 8},  1653 (2011).




\end{thebibliography}
\end{document}